# Non-thermal breaking of magnetic order via photo-generated spin defects in the spin-orbit coupled insulator $Sr_3Ir_2O_7$

*Ernest Pastor[1†\*], David Moreno-Mencía[1†], Maurizio Monti[2], Allan S. Johnson[1], Nina Fleischmann[1], Cuixiang Wang[3], Youguo Shi[3], Xuerong Liu[4], Daniel G. Mazzone[5], Mark P.M. Dean[6], Simon Wall[\*1,2]*

[1]ICFO-Institut de Ciències Fotòniques, The Barcelona Institute of Science and Technology, 08860, Castelldefels, Barcelona, Spain.
[2]Department of Physics and Astronomy, Aarhus University, Ny Munkegade 120, 8000 Aarhus C, Denmark.
[3]Beijing National Laboratory for Condensed Matter Physics, Institute of Physics, Chinese Academy of Sciences, Beijing 100190, China
[4]School of Physical Science and Technology, ShanghaiTech University, Shanghai 201210, China.
[5]Laboratory for Neutron Scattering and Imaging, Paul Scherrer Institut, CH-5232 Villigen, Switzerland
[6] Condensed Matter Physics and Materials Science Department, Brookhaven National Laboratory, Upton, New York 11973, USA.
[†]Equal contribution

*Corresponding authors: ernest.pastor@icfo.eu, simon.wall@phys.au.dk*

**Abstract**: In many strongly correlated insulators, antiferromagnetic order competes with exotic and technologically relevant phases, like superconductivity. While control of the spin order is critical to stabilize new functional states, elucidating the mechanism of laser-induced demagnetization in complex oxides remains a challenge. In particular, it is unknown if the optical pulse can quench magnetization non-thermally or if it instead only acts as a heat source. Here we use ultrafast, broadband, optical spectroscopy to track the response of the electronic, lattice and spin degrees of freedom and its relation to antiferromagnetism in the strongly spin-orbit coupled insulator $Sr_3Ir_2O_7$. We find that magnetization can be rapidly and strongly suppressed on a sub 150 fs timescale. At low excitation fluences the magnetic recovery is fast however, the recovery time increases dramatically with the magnitude of demagnetization. At the same time, we show that the lattice, evidenced through the $A_g$ phonon frequencies, appears to remain below $T_N$, suggesting that the system remains non-thermal during the optical modulation of spin order. We suggest that photo-generated spin defects are responsible for the long-lived demagnetized state and discuss its implications for optical control of solids.

# 1. Introduction

In Mott insulators, the breakdown of long-range antiferromagnetic spin order into phases such as high-temperature superconductivity through doping is a key method to control quantum materials in equilibrium [1,2]. Out of equilibrium, femtosecond light pulses have emerged as a powerful tool to access new non-equilibrium phases, but the role played by the spin degree of freedom remain poorly understood. In particular, it remains unknown if optical quenching of the spin order requires a concomitant heating of the lattice, i.e. if the suppression of spin order is due to rapid, but trivial, heating, or if in contrast, the suppression is non-thermal, potentially enabling access to new transient states unreachable in thermodynamic equilibrium. [3] Central to understanding this issue is understanding how photo-excitation influences all degrees of freedom in the system. However, tracking many different degrees of freedom often requires the use of multiple techniques, sometimes operating in very different regimes [4] and such an approach can lead to artefacts. [5]

Ultrafast optical spectroscopy is a powerful method for tracking structural and electronic transitions. Structural changes can be inferred from the dynamics of the coherent phonon response [6,7] while electronic changes can be modelled through changes in the optical conductivity [8]. However, finding unique optical markers for antiferromagnetism is particularly challenging as the lack of net magnetic moment hinders the use of traditional magneto-optical methodologies. Such difficultly has stimulated the development of new methods to probe it on fast timescales, but many require access to large scale user facilities that makes collecting full datasets difficult and often sensitivity to magnetic order comes at the expense of sensitivity to other degrees of freedom. [2,9–16]

Here, we demonstrate that broadband "4D" optical spectroscopy can extract quantitative information on the electronic, structural *and* magnetic degrees of freedom in an antiferromagnetic insulator. We record the transient optical response of a correlated antiferromagnetic insulator in four dimensions namely, as a function of time, wavelength, excitation fluence and temperature. This multidimensional approach allows us to separate out magnetic dynamics from structural and electronic processes enabling us to map out quantitative magnetic changes, while also assessing the lattice temperature from the viewpoint of the phonon frequencies. We show that in this system, moderate photo-excitations causes a large, ~50%, and prompt suppression of magnetic order which can recover rapidly (under a picosecond). In contrast, at higher excitation densities, the magnetic order can be completely suppressed, resulting in a dramatic increase of the recovery time. Crucially, we demonstrate that the lattice temperature remains below the equilibrium Néel temperature ($T_N$) while the system is in the demagnetized state, showing that the long-lived state is non-thermal. Our experiments and simulations suggest that the formation of photo-induced spin defects dictates the stability of the demagnetized state and the recovery of magnetic order. The picture that emerges from our data is one in which the degree of the system's disorder plays a dominant role in determining the fate of the photo-excited state offering an opportunity to tune antiferromagnetic order without changing the temperature.

We focus on the Ruddlesden-Popper family of iridium- oxides with formula $Sr_{n+1}Ir_nO_{3n+1}$ in which *n* indicates the number of $SrIrO_3$ perovskite layers located in between SrO layers. [17,18] These oxides have raised much interest due to their similarities with superconducting cuprates and because their strong spin-orbit coupling generates a $J_{eff}=1/2$ insulating state (**Figure 1B**) that can lead to new functionality. [19–21] Herein, we evaluate the bilayer (n=2) system, $Sr_3Ir_2O_7$, which displays an out-of-plane-collinear antiferromagnetic insulating state that emerges from a paramagnetic insulating state at the Néel temperature, $T_N$~280 K. [22] Unlike $Sr_2IrO_4$ (n=1), the spin-wave dispersion in $Sr_3Ir_2O_7$ is gapped. As a result, high-energy modes cannot decay through coupling to lower energy modes and decay can only occur through coupling to other degrees of freedom. This can slow thermalization of the spin system with other degrees of freedom and may enable long-lived non-thermal states. Indeed, time-resolved resonant inelastic X-ray scattering measurements of $Sr_3Ir_2O_7$ have shown spin excitations throughout the magnetic Brillouin Zone [20] which are longer-lived than their counterparts in $Sr_2IrO_4$. [9] However, the non-thermal nature of the transient state could not be verified. In the following, we show how broadband optical spectroscopy can complete this goal and shed light into the demagnetization mechanism.

Our paper is structured as follows, in section 2 we discuss the origin of the static optical properties of $Sr_3Ir_2O_7$ as well as how the optical region is sensitive to the magnetic order and how photoexcitation changes the optical spectra both above and below the Néel temperature. In section 3, we show how, through wavelength selectivity, we can isolate the magnetic dynamics from the other degrees of freedom, enabling us to extract detailed information on the magnetic response as a function of fluence, time and temperature. In section 4, we then focus on the response of the lattice obtained from the coherent phonon response and show that, despite a significant thermal red-shift in the phonon mode frequency with temperature, no red-shift is observed following photo-excitation indicating that the lattice is in a non-thermal state. In addition, we discuss how the phonons also influence the electronic properties of the material. Finally, in section 5 we present a discussion of a thermal and non-thermal scenarios for $Sr_3Ir_2O_7$ and present simple Monte Carlo simulation to demonstrate how increasing light-induced disorder can explain the spin dynamics without need for a temperature change.

## 2. Temperature and light-induced changes in reflectivity

We start by evaluating the change in reflectivity as a function of temperature (**Figure 1A**). We focus on the visible region which, as depicted in **Figure 1B**, primarily probes charge transfer transitions involving lower-lying O 2p and Ir 5d states to unoccupied states near the band gap. As the band-gap has a Slater-like character, it is thus expected to be magnetically sensitive [17,23,24], with the long-range magnetic order perturbing the optically-probed magnetic structure. Indeed, we observe that the static reflectivity significantly increases above $T_N$ but the change is gradual and not abrupt at the transition temperature (**Figure 1A**). This likely reflects the fact that the optical changes are not exclusively associated with magnetism but also depend on the electronic and structural degrees of freedom. As these also evolve with temperature, isolating the magnetic signal is non-trivial and hinders accurate quantification of the magnetic order from the reflectivity alone.

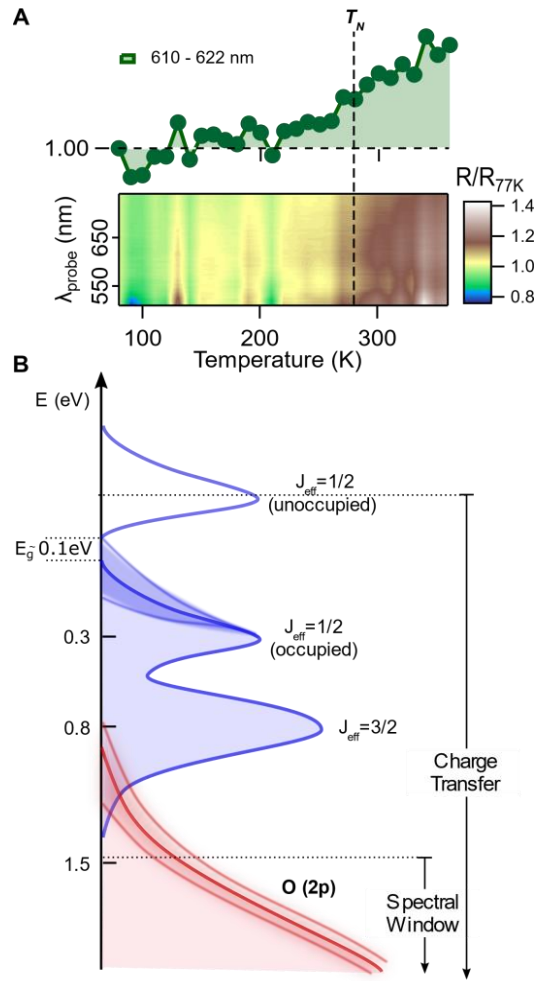

**Figure 1: $Sr_3Ir_2O_7$ and its optical response**. (A) Temperature dependence of the static reflectivity above and below the Néel temperature $T_N$~280 K. The signal is shown relative to the value at 77 K (AFM). Top: Lineout at ~610 nm showing a change in reflectivity between low and high temperatures. For clarity, the signal is smoothed; raw data are displayed in Figure S1. (B) Schematic summary of the electronic structure of $Sr_3Ir_2O_7$. [23] The strong spin orbit coupling generates a $J_{eff}$=1/2 Mott-like insulating state characterized by relatively small bandgap ($E_g$), a filled Lower Hubbard Band (LHB) and an empty Upper Hubbard Band (UHB). The spectral window of our broadband experiments is primarily sensitive to changes in the amplitude and energy of the charge transfer resonance induced by the magnetic, structural and electronic degrees of freedom.

This challenge can, however, be overcome with ultrafast broadband spectroscopy. **Figure 2A** shows typical broadband transient reflectivity spectra taken above and below $T_N$, following photoexcitation with 800 nm (1.5 eV, F = 5 mJ cm$^{-2}$) laser light (other fluences displayed in Figure S2). The spectral changes are qualitatively similar in both cases. Both temperatures show a suppressed reflectivity at short wavelengths and an increase above ~600 nm (see lineouts on the top panel) consistent with a photo-induced shift in the O2p-Ir5d charge transfer resonance (red line in **Figure 1B**), similar to what is seen in the cuprates [25]. Moreover, both temperatures exhibit fast dynamics within the first picosecond, followed by a much slower plateau. The signal is also modulated by an oscillating component due to coherent phonons, which will be discussed in section 4.

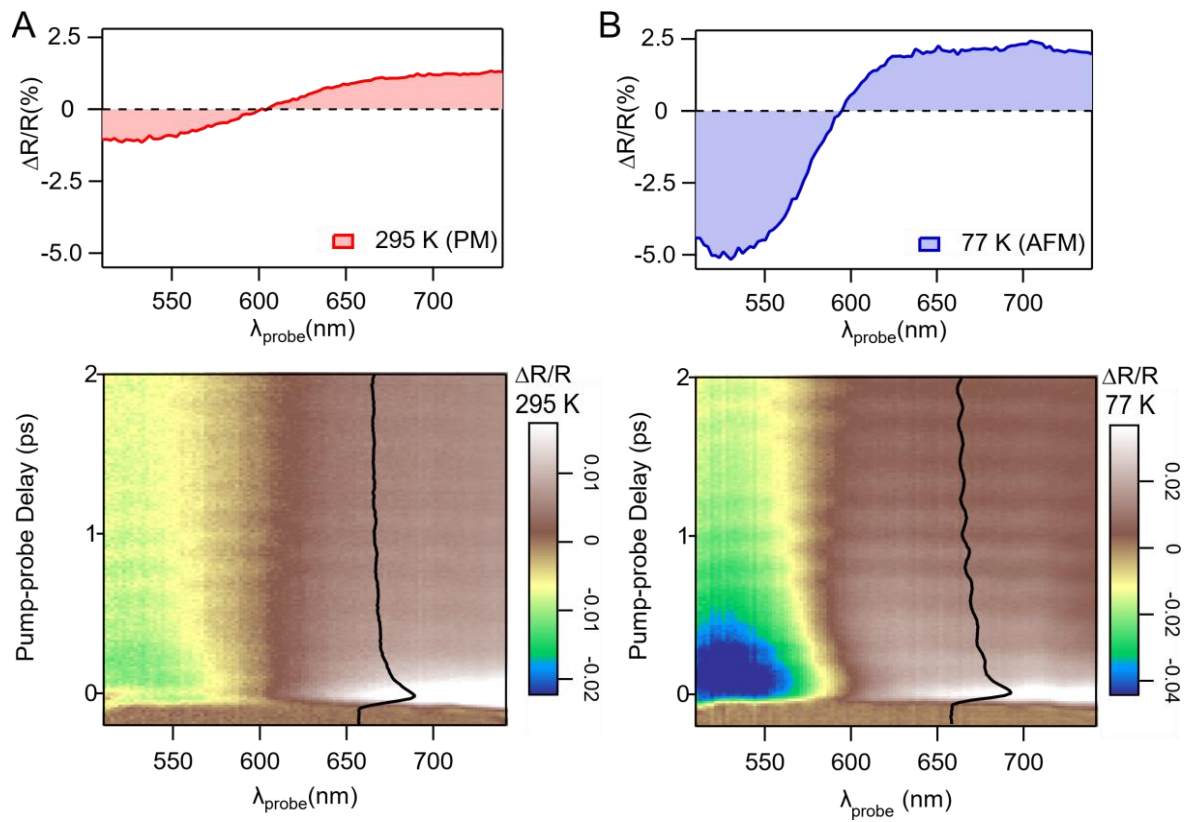

**Figure 2: Transient reflectivity response:** Transient reflectivity change upon 800 nm excitation (F = 5 mJ cm$^{-2}$) at (A) 295 K (paramagnetic, PM) and (B) 77 K (antiferromagnetic, AFM). For each temperature, the top panel shows a lineout at 0.14 ps time delay. Also displayed an example of the coherent oscillations that modulate the transient response (obtained within 650-700 nm range). The amplitude of the oscillations is notably larger at 77 K when the sample is in the antiferromagnetic state (for full kinetic traces see Supporting Information)**.**

At room temperature, where the system has no magnetic order, the dynamics result from purely electronic and structural changes. Below $T_N$, the spectral response is similar, but with a different magnitude. In particular, we observe a significant signal enhancement around 500-650 nm, while the changes above 650 nm are less abrupt. In addition, the kinetic lineouts show a marked increase in the amplitude of the oscillations with decreasing temperature, as discussed later. While the different response between the antiferromagnetic and paramagnetic states indicate a fingerprint of magnetic behavior, it remains unclear whether the observed optical signals can be attributed only to the magnetic degrees of freedom.

### 3. **Optical fingerprints of antiferromagnetic order**

In general, each degree of freedom will produce a different spectral change in the dielectric function, which is the property measured in our experiment. In the linear approximation, one can then express the measured change in the reflectivity due to the changing dielectric function as

$$\delta R = \frac{dR}{d\epsilon}\left(\frac{d\epsilon}{dC}\delta C + \frac{d\epsilon}{dQ}\delta Q + \frac{d\epsilon}{dM}\delta M\right) \quad (1)$$

Where C, Q, M represents the charge, structural and magnetic degrees of freedom respectively. These three terms will have different wavelength dependences, therefore, to establish which optical region has the most sensitivity to magnetic order and least sensitivity to the other degrees of freedom, we measure a 4D dataset, recording the response of the system as a function of probe wavelength, $\lambda$, fluence, $F$, temperature, $T$, and time, $t$. Our aim is to identify a region of the spectra that is most insensitive to thermal changes in the electronic and structural properties above $T_N$. This allows us to subsequently subtract these contributions from the signal below $T_N$ and thus to isolate the magnetic response. **Figure 3A** shows a detailed temperature scan at different probe wavelengths for a fixed fluence and delay, which exemplifies the merits of our approach. From the data in **Figure 2** alone (comparison of spectra at 295 K and 77 K), one might assume that the 500-550 nm region would exhibit the largest magnetic sensitivity as it shows the largest change when crossing $T_N$. However, the full temperature scan reveals that this region is strongly temperature dependent above $T_N$, which precludes extraction of the magnetic signal without a detailed understanding of the non-magnetic temperature evolution of the system. Likewise, at long wavelengths (> 700 nm) the signal shows a temperature independent response above $T_N$, but lacks magnetic sensitivity below $T_N$. Between these two extremes, we find that wavelengths around 610 nm are unique because the response is temperature independent above $T_N$, whilst retaining sensitivity to magnetic order. This remains true for all probed fluences and time delays as shown in Figure S3 and as discussed below.

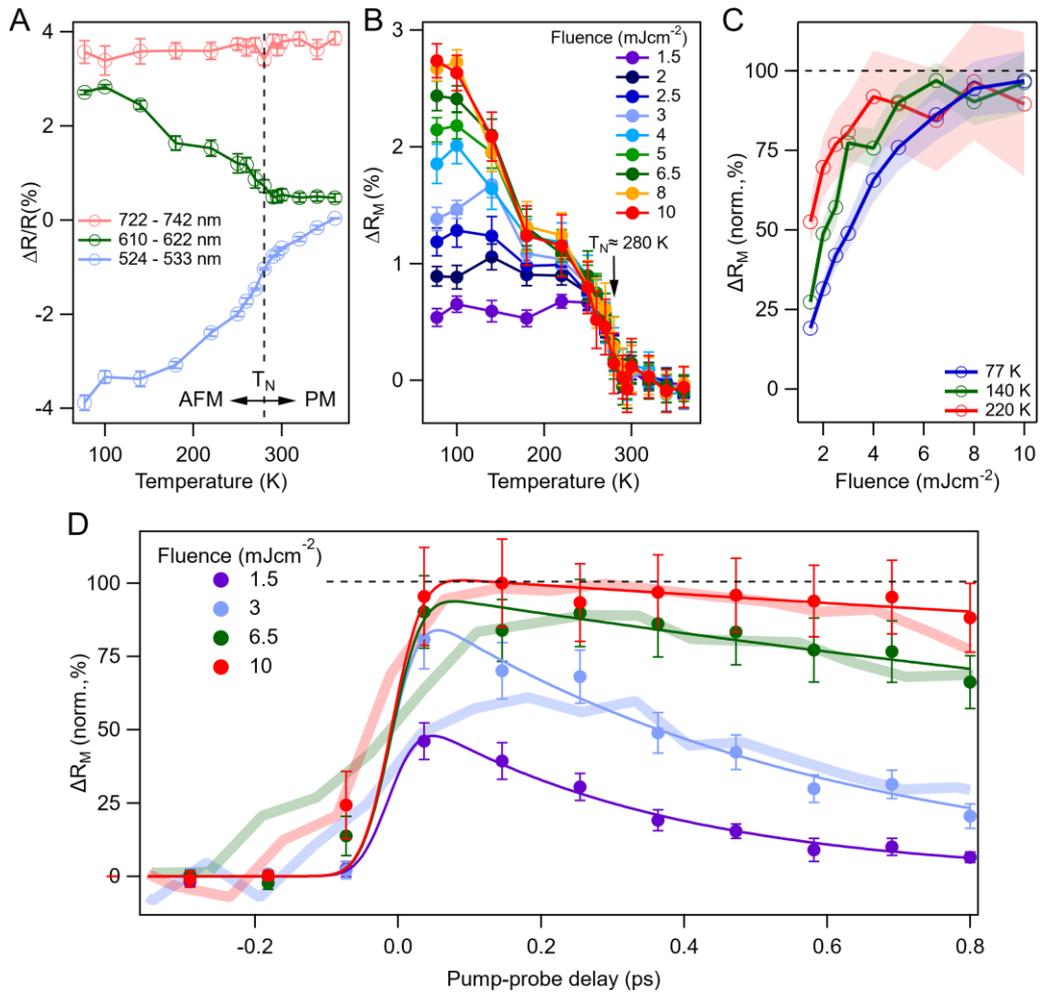

**Figure 3: Changes in the magnetic degree of freedom:** (A) Temperature dependence of the transient reflectivity at different probe wavelengths. For clarity the traces are offset. The 610 nm region is both sensitive to magnetic order while showing a temperature independent response above $T_N$. (B) Temperature dependence of the magnetic degree of freedom ($\Delta R_M$) at different excitation fluences measured at 360 fs time delay obtained as $\Delta R(T) - \Delta R(T>T_N)$ in the 610 nm region. The signal saturates at high fluences and at temperatures close the Néel temperature as expected for light-induced demagnetization. We normalize the maximum $\Delta R_M$ value at 10 mJ cm$^{-2}$ to 100% demagnetization. (C) Fluence dependence of the magnetic signal at different temperatures showing that full demagnetization is more rapidly attained at higher temperatures. (D) Time-evolution of the magnetic signal measured at 77 K at different fluences showing fast and large (>40%) demagnetization as well a change from rapid recovery to permanent demagnetization at F > 4 mJ cm$^{-2}$ (See also Figure S5). For comparison, the time evolution of the magnetic Bragg peak obtained from [20] is also shown (solid translucent lines, scaled).

We explore the feasibility of isolating the magnetic component by subtracting the electronic, temperature independent, contribution from the total reflectivity ($\Delta R(T) -$

ΔR(T>$T_N$)) around 610 nm. We label this assay of magnetism as Δ$R_M$. **Figure 3B** shows this analysis for data collected at 360 fs time delay at different excitation conditions (see Figure S4 for other time delays and details). Above $T_N$ the signal is zero at all fluences, as expected by our methodology. At $T_N$ there is a discontinuity and the magnetic response rises sharply. Such jump with temperature contrasts with the much slower change seen in reflectivity measured statically (**Figure 1A**), and is indicative that our magnetic assay is able to isolate the light-induced change in magnetic order from the other degrees of freedom. Further corroboration of the magnetic sensitivity can be inferred from the fluence dependence of the signal. Increasing the pump fluence increases the signal, indicating that the system shows a greater degree of demagnetization. Moreover, for the highest fluences the signal saturates, as would be expected for a complete demagnetization (**Figure 3C**). Importantly, both the saturation fluence and signal magnitude are temperature dependent. The magnitude of the signal at saturation increases as the sample is cooled below $T_N$; this is indicative of the fact that a greater fraction of the material displays long-range magnetic order at lower temperatures, which can then be melted by the pump pulse and manifest in the optical response. Furthermore, the shift in the saturation fluence with temperature is consistent with the fact that more energy is needed to demagnetize more of the sample. Based on this, we assign the saturation fluence as the fluence required to completely suppress magnetic order. For all temperatures we observe that fluences above ~4-5 mJ cm$^{-2}$ result in saturation. Thus, we normalize Δ$R_M$ to the value obtained at the highest excitation fluence (F = 10 mJ cm$^{-2}$), which provides a scale for a quantitative analysis of the demagnetization. **Figure 3D** shows the change in the normalized magnetic signal at 77 K when the sample is in the antiferromagnetic state (see also Fig S5). We observe a sharp change in the magnetic signal immediately after photoexcitation. Notably, we find a very large (>50%) demagnetization even at low fluences, which is a strong indicator that the magnetic order can be efficiently perturbed by the laser pulse in this material.

We note that our conclusions rely on two assumptions. Firstly, we assume that higher-order cross-couplings between the different degrees of freedom, not considered in equation 1, do not significantly impact the signal. Second, we assume that the high-temperature response still remains constant below $T_N$. Neither case is expected to strictly hold. For example, the phonons de-phase less at lower temperatures. Thus, subtracting the high temperature phonon response from the low temperature data will still show a residual coherent oscillation that should not be attributed to a modulation on the magnetic response. However, we assume that these deviations and others are small compared to the resolution of our data.

In order to confirm the validity of our approach, we overlay to our signal, the time evolution of a magnetic Bragg peak measured at several (high) fluences using XFEL radiation and which is a direct probe of magnetic order [20]. Due to difference in the pump wavelength and differences in penetration depth of the probe, the two data sets need to be scaled in fluence (see Supporting Information), but the agreement between the optical and x-ray data is remarkable and demonstrates that the potential issues with our method of analysis do not have a quantitate influence on our results within our measurement accuracy. The identification of a reliable marker of magnetic order via reflectivity measurements, while not a direct probe of the spin system, provides a tool to systematically map out the response of the sample under a broad range of experimental conditions to those achievable during XFEL measurement.

Inspection of the data in **Figure 3D** reveals that the recovery dynamics of the spin system are strongly fluence dependence. Following the initial rapid demagnetization step, at low fluences the signal decays and recovers rapidly. In contrast as saturation fluence is reached only a minimal decay within 1 ps occurs, indicating near-complete and long-lasting demagnetization. The observed fluence dependence, suggests that the laser excitation is able to trigger different mechanisms of de- and re-magnetization depending on the strength of the perturbation. Such behavior can have important implications for optical control of magnetism. Yet, from these data alone, it is challenging to understand what is driving this change in behavior and measurements of the other degrees of freedom are needed. However, comparing different experimental techniques can be challenging. This is particularly true when comparing excitation fluences often recorded in very different conditions. [5] By validating our optical probe of magnetism, we are now able to confidently compare the response of the charge and structural degrees of freedom with the known suppression of the magnetic order.

In particular, we wish to understand whether the response of the magnetic system can be simply interpreted in terms of a three-temperature model or other mechanisms are at play. In a classical thermal scenario, the photo-excited electrons rapid heat the spin system above $T_N$ triggering the loss of long-range order. Subsequently, the spins cool by transferring energy to the lattice. After spin-lattice thermalization, if the lattice temperature remains below $T_N$, the magnetic order should recover. In contrast, if the lattice temperature rises above $T_N$ the system remains in a demagnetized state until the lattice temperature cools via thermal diffusion, which is generally slow. Such behavior has been previously observed in metallic systems, [3] and points toward strategies of optical control in which the laser acts effectively as a heat source, but may not be applicable in systems with energy gaps that can prevent thermalization. Alternatively, the magnetization dynamics may be non-thermal and follow a different path in which the electronic, spin and lattice degrees of freedom are not in thermal equilibrium, thus providing exotic paths for optical control. To understand which model applies, it is essential to establish the response of the lattice and importantly, with optical spectroscopy, these details can be obtained in the same experimental conditions.

### 4. <u>Tracking the lattice degree of freedom</u>

In order to explore the role of the lattice during the magnetic dynamics, we take advantage of the coherent oscillations that modulate the transient reflectivity. Analysis of the high time-resolution scans were performed at room temperature and at 77 K and under low fluence excitation, yield two frequencies at ~146 cm$^{-1}$ (4.4 THz) and

~182 cm$^{-1}$ (5.4 THz) corresponding to A$_{1g}$ modes (see **Figure 4A** and supplementary materials for details). As expected, these modes red shift at higher temperature with a concomitant decrease in amplitude. The values obtained are in good agreement with those measured with equilibrium Raman spectroscopy [22]. We can then use this temperature dependence of the mode as a proxy for the lattice temperature as we increase the excitation fluence. **Figure 4B** shows that the frequency of both A$_{1g}$ modes only exhibit a small red shift with fluence, which is less than that the room temperature value. This strongly points to the fact that even at high fluences, where magnetism is completely suppressed for multiple picoseconds [20], the lattice heats, but remains below T$_N$.

Further evidence for a non-thermal state of the system can be found by looking at how the phonon dynamics influence the electronic states. **Figure 4C** shows the wavelength dependence of the amplitude of each phonon mode at 77 K at different fluences. We observe that the 182 cm$^{-1}$ mode couples almost exclusively to short wavelengths whereas the 146 cm$^{-1}$ mode is boarder and primarily couples to longer wavelengths. Most strikingly, we find that the peak of the 146 cm$^{-1}$ mode's electronic resonance red shifts with increasing excitation, while the other mode remains unchanged. Beyond these spectral changes, we find that the amplitude of the high frequency mode increases slowly in a linear-like fashion with excitation fluence. In contrast, the low-frequency mode increases linearly and rapidly up to ~4-5 mJcm$^{-2}$ and subsequently plateaus and exhibits a slight decrease at the highest fluences (**Figure 4D**). The red shift in the wavelength maxima is an indicator that the 146 cm$^{-1}$ mode shifts the charge transfer resonance that dominates the spectral window, as previously observed for the analogous mode in bilayer cuprates. [25] The mode amplitude saturation at moderate fluences is in good agreement with the saturation of the magnetic signal observed in **Figure 3B** and may suggest a strong link of the 146 cm$^{-1}$ mode in particular with the magnetic order of the system. We note that the fluence at which the mode's amplitude decreases matches the fluence at which the mode's frequency starts to shift; this could be the result of more significant lattice heating in the high photoexcitation regime. Additional corroboration of the sensitivity of the modes to magnetic order can be found in their spectral response at different temperatures. We observe that above $T_N$, at 295 K, the spectral dependence of the ~182 cm$^{-1}$ vibration is the same as at 77 K, while, the ~146 cm$^{-1}$ mode becomes broad and featureless (Figure S6), further indicating its high sensitivity to magnetization. However, from optical measurements alone, we cannot determine if the amplitude of the atomic displacements is saturating, or if it the coupling of the mode to the electronic states is changing.

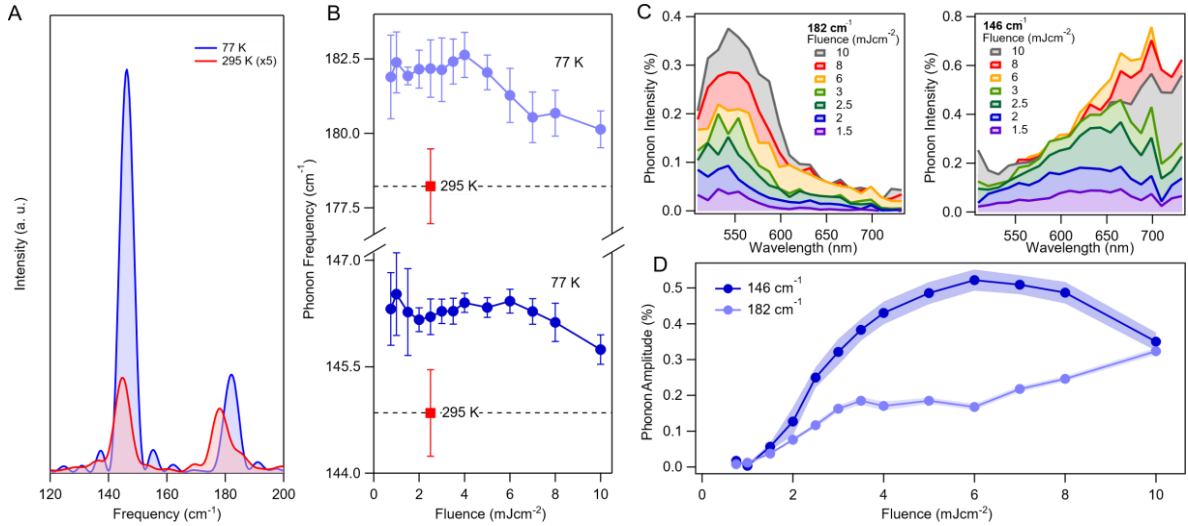

**Figure 4: Changes in the lattice degree of freedom** (A) Fourier transform of the oscillations in the transient reflectivity at 77 K (AFM) and 295 K (PM) at F = 2.5 mJcm$^{-2}$. The characteristic high and low energy $A_{1g}$ modes can be captured. The amplitude decreases and the central frequency shifts with increasing temperature. (B) Change in the mode frequency as a function of excitation fluence at 77 K (AFM). For comparison the frequency measured at 295K (PM) at low fluences is also displayed (red square). Despite large demagnetization achieved at all fluences, the modes remain mostly unchanged and exhibit only an inflection at the highest fluences. (C) Spectral response of the two $A_{1g}$ modes as a function of fluence. The 182 cm$^{-1}$ mode couples to low wavelengths while the 146 cm$^{-1}$ mode couples to longer wavelengths and exhibits a red shift with increasing fluence. (D) Maximum amplitude of the phonon modes as a function of fluence (obtained by fitting the spectral response, see SI for details).

5. **Discussion**

We have used broadband optical spectroscopy to identify markers for both the spin and lattice degree of freedom. In agreement with XFEL measurements, we observe that laser excitation can induce large (>50%) changes in antiferromagnetic order even at low fluencies. However, the magnetization dynamics are strongly fluence dependent. Below approximately 4-5mJcm$^{-2}$ the system is rapidly demagnetized and quickly restored. In contrast, at higher fluences the demagnetization saturates and becomes longer-lived. Moreover, by tracking the coherent phonons in the system we observe that the low frequency $A_{1g}$ mode frequency is only moderately perturbed and remains consistent with a low temperature lattice, despite the complete melting of magnetic order. In addition, the coupling between the phonon mode and the electronic structure also reports evidence of the demagnetized state.

These results strongly suggest that the magnetic system shows a non-thermal response and the long-lived demagnetized state is not simply a state in which the lattice temperature has risen above $T_N$. Hence our data rules out multi-temperature models [3] and indicates that lattice temperature increases are not required to either suppress long-range spin order, nor prevent its recovery. Instead, our observations can be explained by considering how photo-excitation modifies magnetic order in insulating quantum materials with strong spin-orbit coupling (SOC). Optical excitation creates electron-hole pairs causing changes in orbital occupancies. Due to the small charge gap in $Sr_3Ir_2O_7$, it is energetically favorable to create zero-spin doubly occupied sites following photoexcitation effectively generating 'spin defects' in the system (**Figure 5A**). [21] This configuration is unstable and at low excitation density the excited electron can recombine with its corresponding hole via the emission of phonons. Although it is in principle possible for any electron to recombine with any available hole (leading for example to hole motion), the strong exchange coupling, which is only slightly smaller than the charge gap, disfavors an electron from recombining with a hole that does not result in antiparallel spin alignment. This is depicted in **Figure 5A-B**. If the electron-hole pair recombine before the surrounding spin system has time to react, the magnetic order can be restored without disordering the spin system beyond the initial creation of the spin defects. This suggests a direct link between electrons rapidly relaxing back to the valence band and the fast recovery of magnetic order, as measured at low excitation fluences. We indeed observe that the magnetic recovery in such a regime and the electronic relaxation times measured above $T_N$ are comparable, suggesting electron-hole recombination in the low-density limit can restore the initial spin configuration. Notably, this process is markedly non-thermal and represents magnetic perturbations beyond linear spin-wave theory. At higher excitation fluences, a large density of spin defects is created, as depicted in **Figure 5C**. When these defects are close, the energy barrier preventing electron-hole recombination events that result in non-antiparallel spin alignments is significantly reduced. This result in a mixture of parallel and antiparallel spin alignments that eventually disorders the underlying spin network, slowing down the regeneration of the long-range order.

To test the feasibility of this scenario, we examine the recovery of a 3D spin-1/2 Ising model following a change in the magnetic order using a Monte Carlo simulation. This model neglects the specific details of $Sr_3Ir_2O_7$, but instead focuses on how order is established from various disordered states while the bath temperature is kept constant. We assume the creation of doubly occupied sites can be simplified as direct spin flips, with the flipping fraction depending on the laser fluence, but no temperature change occurs. We then count how many Monte-Carlo steps are required to return to the equilibrium conditions (see supplementary information for details on the Monte-Carlo simulation). **Figure 5D** shows that when the spin network is only weakly perturbed, order recovers quickly. However, when the system approaches complete demagnetization, the recovery slows down dramatically and the system remains demagnetized, despite the temperature of the simulation being significantly below $T_N$ (see also Figure S8 which includes simulations which increase the lattice temperature). This fast recovery at small perturbations and the generation of a long-lived state at high defect concentrations in the absence of system heating is in good agreement with our experimental observations. We emphasize that this basic model does not honor the complexity of $Sr_3Ir_2O_7$. For example, the model is blind to the mechanism of perturbation beyond imposing an instantaneous change, which is something that will depend on the dominant energy scale of the system. Indeed, while charge transfer in any Mott insulator is expected to produce rapid demagnetization [26], the exact nature of the spin-orbit coupling and the exchange interaction are likely key factors for explaining the magnitude and speed of the switching process [27]. Notwithstanding the limitations of the model, this approach allows us to explore the initial basic response of a system to a disruption of long-range order and highlights the role played by the degeneracy of the spin-disordered state in preventing a rapid recovery. This points towards a central role of disorder in determining the physics of irradiated solids [28,29] and may be best understood as glass-like systems. [30] While disorder has been largely regarded as undesirable, increasing evidence suggest it might be essential in enabling and controlling functionality in systems such as solar cells and photocatalysts and might provide new paths to control the excited state [31–33].

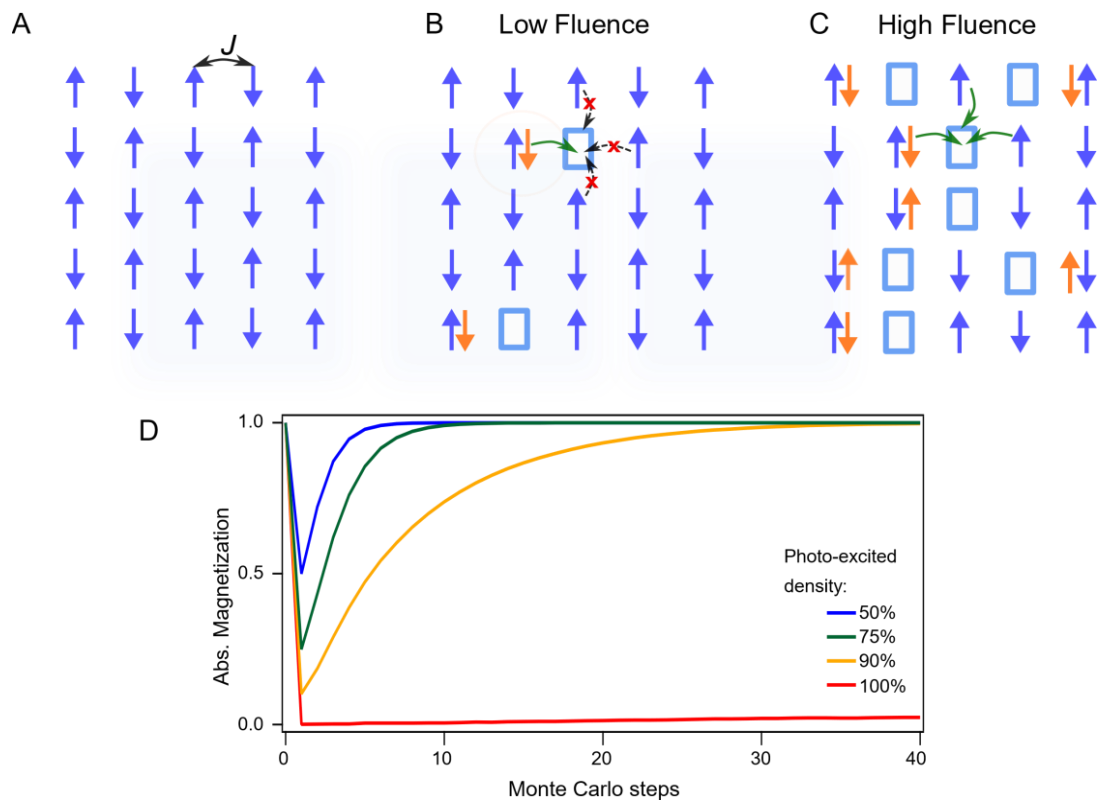

**Figure 5: Disorder mediated recovery of magnetic order.** (A) Ground state characterized by a large exchange coupling (*J*) that leads to antiferromagnetic (AFM) order. (B) Photoexcitation at low fluences results in the formation of a small density of spin vacancies and doubly occupied states with zero net moment. Effective destruction of the moment is enabled by the small charge gap. Rapid recombination events regenerate the antiferromagnetic order but recombination paths that produce parallel spin alignments are prevented by the dominant antiferromagnetic exchange coupling (C) Photoexcitation at high fluences. The formation of a large density of spin vacancies relaxes the recombination constraint leading to a long-lived disordering of the spin system. (D) Monte Carlo Magnetization dynamics in a 3D spin-1/2 Ising model resulting from an instantaneous and random spin flips at different degrees of excitation. If a small fraction of spins is flipped the system recovers fast. In contrast when a large number of spins are flipped inducing large disorder, long-lived change in magnetization occurs.

Another key observation in our data is that the electronic coupling to the low-frequency $A_{1g}$ mode is strongly modulated by temperature and photoexcitation and, correlates with changes in the magnetic order. In particular, we observe that the mode spectrum is highly temperature sensitive and red-shifts with increasing fluence. Moreover, the amplitude of the mode increases linearly with excitation and saturates at a fluence coinciding with the full demagnetization of the sample. In contrast, the optical window modulated by the high frequency $A_{1g}$ mode does not strongly change with temperature

Page **14** of **32**

and fluence. We note, there is no consensus in the literature on the assignment of the $A_{1g}$ vibration in $Sr_3Ir_2O_7$. However, in analogy to bilayer cuprates of the similar structure, it is likely that the mode is responsible for the degree of charge transfer between the metal and the oxygen centres within the perovskite layers. This is in agreement with the strong coupling of the mode to visible wavelengths, which probe the charge transfer resonance. Activation of this mode has strong implications for functionality control as, for example, in cuprates the charge transfer is known to correlate with the superconductive state. [25] Specifically for iridates, in our experiments, charge excitation shifts the phonon equilibrium position and quenches magnetization depending on the density of spin vacancies generated. Direct perturbation of the mode could provide a pathway to modulate both the degree of spin disorder and the exchange interaction in the system and ultimately control antiferromagnetic order.

## 6. Conclusion

In summary, although optical reflectivity does not couple to magnetism directly, we have outlined a strategy to infer magnetic behavior from the wavelength-dependent optical reflectivity. Using this marker, we have shown how photo-excitation of $Sr_3Ir_2O_7$ can non-thermally suppress the magnetic long-range order. We have found a regime, at low fluences and low temperatures, in which we are able to manipulate the magnetic order by up to 50%, but where the system can recover within a picosecond. In contrast, at higher fluences we measure a crossover to a regime in which the magnetization is completely suppressed, but where the lattice temperature remains below $T_N$. We argue that such crossover emanates from the generation of photoinduced spin defects that enable a pathway for spin disordering and long-lived breaking of magnetic order without significant lattice heating. We have also identified a Raman active mode that is strongly coupled to the electronic structure of the magnetic state. The light-induced, non-thermal change of spin order opens new opportunities to investigate transient phases in solids. In particular, the iridates have long been thought of as the 5d analogues of cuprates, but the disruption of magnetic order through static doping did not allow to induce superconductivity. Optical excitation and manipulation of long-lived spin order via charge or through phonon control [34,35] might enable a new route to access and stabilize superconductivity.

## Acknowledgements

This work was funded through the European Research Council (ERC) under the European Union's Horizon 2020 Research and Innovation Programme (Grant Agreement No. 758461), by PGC2018-097027-B-I00 project funded by MCIN/ AEI /10.13039/501100011033/ FEDER "A way to make Europe", the Spanish State Research Agency through the "Severo Ochoa" program for Centers of Excellence in R&D (CEX2019-000910-S), the Fundació Cellex, and Fundació Mir-Puig, the Generalitat de Catalunya through the CERCA program. A.S.J acknowledge support from the Marie Skłodowska-Curie co-fund a PROBIST fellowship (grant agreement

No. 754510). E.P acknowledges the support form IJC2018-037384-I funded by MCIN/AEI /10.13039/501100011033. X.L. acknowledges support from the National Natural Science Foundation of China under grant No. 11934017. Work at Brookhaven National Laboratory was supported by the U.S. Department of Energy, Office of Science, Office of Basic Energy Sciences under Contract No. DE-SC0012704. C.W. and Y. S. acknowledge the support from the National Natural Science Foundation of China (Grants No. U2032204) and the K. C. Wong Education Foundation (GJTD-2018-01).

# Supplementary Materials

## 1. Materials and Methods

**Materials:** $Sr_3Ir_2O_7$ single crystals were synthesized using the self-flux flux method as described in Reference [1] and references therein.

**Ultrafast Transient Reflectivity measurements:** (Figure 2, main text) were performed using a Legend Elite amplifier (Coherent) with 800 nm central wavelength, 40 nm bandwidth a pulse duration of ~35 fs with and an energy of 5 W at 5 kHz repetition rate. The broadband visible probe light was generated by focusing a small fraction of the 800 nm beam onto a sapphire crystal while 800 nm light was mechanically chopped at 2.5 kHz and was used as the pump pulse. The pump energy was adjusted using a half wave plate on a motorized stage capable of rotating the polarization of the beam relative to a Brewster polarizer. Both beams were focused on the sample and to ensure probing of a homogeneously pumped region, the probe size (~ $4 \cdot 10^{-5}$ $cm^2$) was made smaller than the pump size (~ $2 \cdot 10^{-4}$ $cm^2$). The changes in the reflectivity of the sample were detected using a commercial Andor Zyla sCMOS camera. The cryogenic measurements were performed using liquid nitrogen cryostat (Oxford Instruments, Optistat DN-V).

**Static Reflectivity measurements** (Figure 1A, main text) were performed on the same sample used in the pump-probe measurements in the cryostat that was also used for the transient reflectivity measurements. A thermal Tungsten-Halogen lamp (Thorlabs SLS201L) was coupled to a multimode fibre and the output imaged onto the sample using a 4f configuration, with a sub-millimeter spot size and an angle of incidence of a few degrees. The reflected light was collected using a second 4f-lens configuration onto a multimode fibre. A cosine corrector was used to reduce the effect of sample drift on the coupling efficiency, and a camera was used to log the sample position in real time to monitor thermal drift. The final spectrum was collected using a silicon CCD spectrometer with 5s exposures (Avantes AvaSpec-Mini4096CL-UVI10). For clarity Figure 1A (main text) displays the measured reflectivity relative to the value at 77 K.

## 2. Analysis of coherent phonons

The transient reflectivity data shown in Figure 2 (main text) is highly modulated at all probe wavelengths by large amplitude oscillations associated with coherent phonons generated by the pump excitation. To extract the frequency of the oscillations the transient data was first differentiated in order to remove the incoherent exponential decay component. Subsequently a fast Fourier transform (FFT) of the resulting differential map yielded a 2D plot of frequency vs probe wavelength. To avoid any coherent artefact at t ~ 0 ps, the dataset was analysed only between 0.3 – 5.5 ps. To obtain a higher signal to noise level, the frequency map was average at all wavelength yielded modes at ~146 $cm^{-1}$ and ~182 $cm^{-1}$ (Figure 4A, main text). The values obtained are in good agreement with those measured with equilibrium Raman spectroscopy [2] and previous ultrafast optical spectroscopy measurements [3]. In order to obtain the frequency value at each fluence (Figure 4B, main text), the differential map at each wavelength was first fitted by the sum of two Gaussians giving by the expression

$$f(x) = A_1 e^{-\left(\frac{x-x_{01}}{w_1}\right)^2} + A_2 e^{-\left(\frac{x-x_{02}}{w_2}\right)^2}$$

Where $A_i$, $x_{0i}$, and $w_i$ (i=1,2) are the amplitude, central frequency and width of the modes. Subsequently, we calculate the average value of the central frequency ($x_{0i}$) of each Gaussian in the probe wavelength range which yields the frequency value shown in Figure 4B (main text). The error bars correspond to the standard deviation of the average calculation for each fluence. Figure S9 shows the different steps of this analysis performed at 77 K and 295 K.

An alternative way to verify such strong fluence dependence of the ~146 cm$^{-1}$ mode at 77 K is by directly fitting the differentiated transient data at different fluence focusing on the 650-700 nm region where this mode shows its maximum intensity and the ~182 cm$^{-1}$ mode contribute less. The differentiated transient data was fit with the equation

$$f(t) = A\cos\left(\frac{2\pi f_1}{T_1} + \varphi_1\right)e^{-t/\tau_1} + B\cos\left(\frac{2\pi f_2}{T_2} + \varphi_2\right)e^{-t/\tau_2}$$

where A, $f_1$, $T_1$, $\varphi_1$ and $\tau_1$ correspond to the amplitude, frequency, period, phase and constant time of the low energy mode respectively. Analogously, the second term of the equation correspond to the high energy mode.

Figure S7 A shows the perfect match between the raw data (circles) and the fit (solid lines) at different fluence values at 77 K. Figure S7 B shows the amplitude parameter from the fit. It shows a similar behaviour on both modes with fluence as observed in Figure 4D (main text).

3. **Calculation of the magnetic contribution**

The transient reflectivity at different delay times (t), probe wavelengths (λ) and fluences (F) shows changes at $T_N$ as $\Delta R = \Delta R(T, \lambda, t, F) = \Delta R_e + \Delta R_M$ where $\Delta R_M$ and $\Delta R_e$ are the magnetic and non-magnetic (e.g. electronic) contributions respectively. A such, $\Delta R_M = 0$ for $T > T_N$. Therefore, $\Delta R_M$ can be obtained provided that $\Delta R_e$ is known. Ideally, the transient response of the electronic system would be a separable function of the different variables (T, λ, t, F), i.e. $\Delta R_e = f(T)g(\lambda)h(t)k(F)$ allowing the functions $g(\lambda)h(t)k(F)$ to be obtained from the high temperature data (>$T_N$). In such case the behaviour of the electron system, f(T), could be obtained below $T_N$. However, we were unable to reliably separate the dataset into functions of specific variables, and no unique spectral function for the electronic degrees of freedom could be found. Therefore, we bypassed this issue by identifying those wavelengths ($\lambda_{e0}$) at which $(d\Delta R_e)/dT(\lambda_{e0})=0$. We found that that the region around $\lambda_{e0}\approx 600$ nm fulfilled this condition. Note that $\Delta R_e$ is not zero, nor is it a constant as a function of time (Figure S3). We assume that the above criteria remains true at this wavelength below $T_N$ and obtain the magnetic contribution as $\Delta R_M = \Delta R(T<T_N, \lambda_{e0}, t, F) - \Delta R(T > T_N, \lambda_{e0}, t, F)$ as shown in Figure S4 and Figure 3 (main text). The validity of this approach is further corroborated by comparison of the optically extracted $\Delta R_M$ with the magnetic Bragg peak obtained from reference [4].

## 4. Comparison of XFEL and Optical data

The X-ray data used in Figure 3D (main text) and reported in reference [4] was performed under grazing incidence and with a 2 μm pump wavelength, whereas the optical work presented herein is performed at 800 nm and a 45 degree angle of incidence. Differences in absorption and reflection coefficients as well as penetration depth miss-matches prevents directly comparing the amplitude of the change and fluence values. In addition, the base temperature of the XFEL data is known with less precision due to the use of a cryo-stream cooling system as opposed to the liquid nitrogen cryostat employed in the optical measurements that allows for fine, controlled changes in temperature. Furthermore, the Magnetic Bragg peak data saturates at a value of ~70% due to the pump probe miss-match between laser and X-rays. This difference should be much smaller at visible wavelengths. Therefore, to correct for this, we scale the magnitude of the X-ray data by a constant factor.

## 5. Details of the Monte Carlo Calculations

Monte Carlo (MC) algorithms and models are commonly used to study equilibrium processes in a variety of systems in physics and beyond. However, they can also be applied to the investigation of dynamical processes [5,6]. Although it is common to implement more complex models [7], in this work we opt for the simplest possible description of an antiferromagnet: the 3D spin-1/2 Ising model with nearest neighbours interaction. We find that this model, in its simplicity, is capable of capturing the essence of the magnetization dynamics after photoexcitation and to qualitatively reproduce the experimental observations.

We model the generation of doubly occupied spin states upon laser excitation as direct spin flips. The modeled system consisted of a 3D simple cubic lattice of lateral size L=100 spins (N=1000000 spins in total). A MC step (MCS) is defined as N attempted spin flips for a system of size N. The system was prepared in a perfectly aligned low temperature state and suddenly disordered at time t=2 MCS by flipping a given percentage of the spins (a percentage calculated from the desired demagnetization value) following a random distribution. After that, the system was let free to evolve for a given number of MC steps. The equilibration was implemented using a Metropolis-Hastings algorithm with Glauber dynamics. At each step the staggered magnetization was measured as the difference of the magnetizations of the two sublattices, which were measured as the sum of all the spins alignment in the sublattice.

In the Monte Carlo dynamics, energy is not conserved. Optical excitation, through spin-flips, increases the energy in the system, but as the magnetic order recovers, the total energy of the system decreases. In a real system this energy would translate to an increase in the temperature. As the probability for spin flips increases with increasing Monte Carlo temperature, it may be expected that such a temperature rise may result in more rapid dynamics as the system can more easily leave local minima. Therefore, a second simulation was performed including an increase in temperature together with the demagnetization. The amplitude of the increase in temperature is proportional to the change in the magnetization. The temperature profile is described by:

$$T = T_O + M \cdot C \cdot (1 - e^{-\frac{t}{\tau}})$$

Where $T_0 = 0.33 T_N$ is the initial temperature, M is the change in magnetization, $C/T_N = 0.56$ is a proportionality constant, which defines the maximum temperature rise relative to a given demagnetization and can be thought of as a heat capacity which converts the non-thermal demagnetization into a temperature. The choice of C=0.56 ensures that the temperature increases substantially above the initial temperature while remaining lower that the transition temperature. This corresponds to a final temperature of $0.89\ T_N$ which is just below the value at which the phase transition starts to affect the dynamics. $t$ is the time in MCS and $\tau$ is the characteristic time of the increase in temperature. $\tau$ is set to be 1MCS to roughly match the fast time constant in the low fluence simulation. However, the resulting dynamics did not change if the temperature rise was considered to be a step function in time. The top panel of Figure S8 reports the result of such simulation along with the corresponding temperature profile in the bottom panel. The presence of the temperature change does not affect the magnetization dynamics which looks very similar to those reported in Figure 5 in the main text as long as the lattice temperature remained below $T_N$.

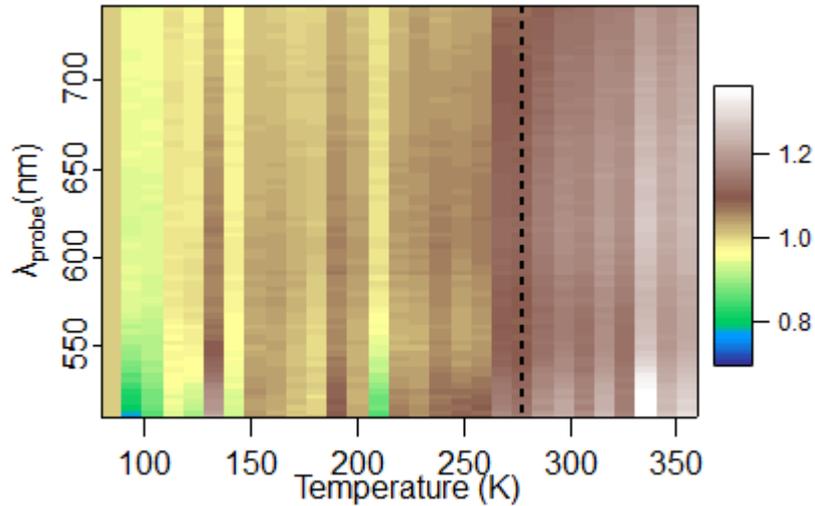

**Figure S1:** Raw temperature dependence of the static reflectivity which was smoothed in Figure 1A of the main text. The signal is shown relative to the value at 77 K (AFM).

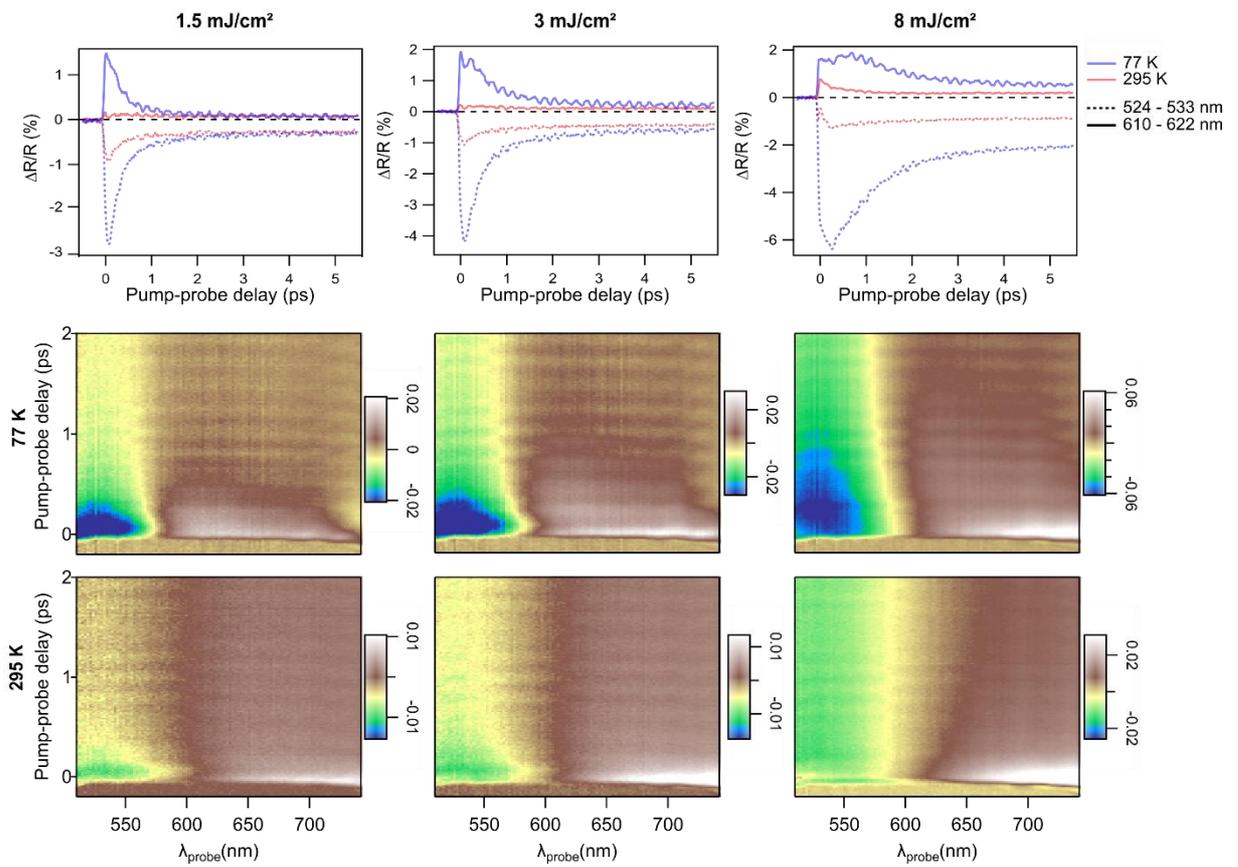

**Figure S2:** Transient Reflectivity change upon 800nm excitation at 77 K (top) and 295 K (bottom) measured at $F$ = 1.5 mJ cm$^{-2}$, 3 mJ cm$^{-2}$ and 8 mJ cm$^{-2}$.

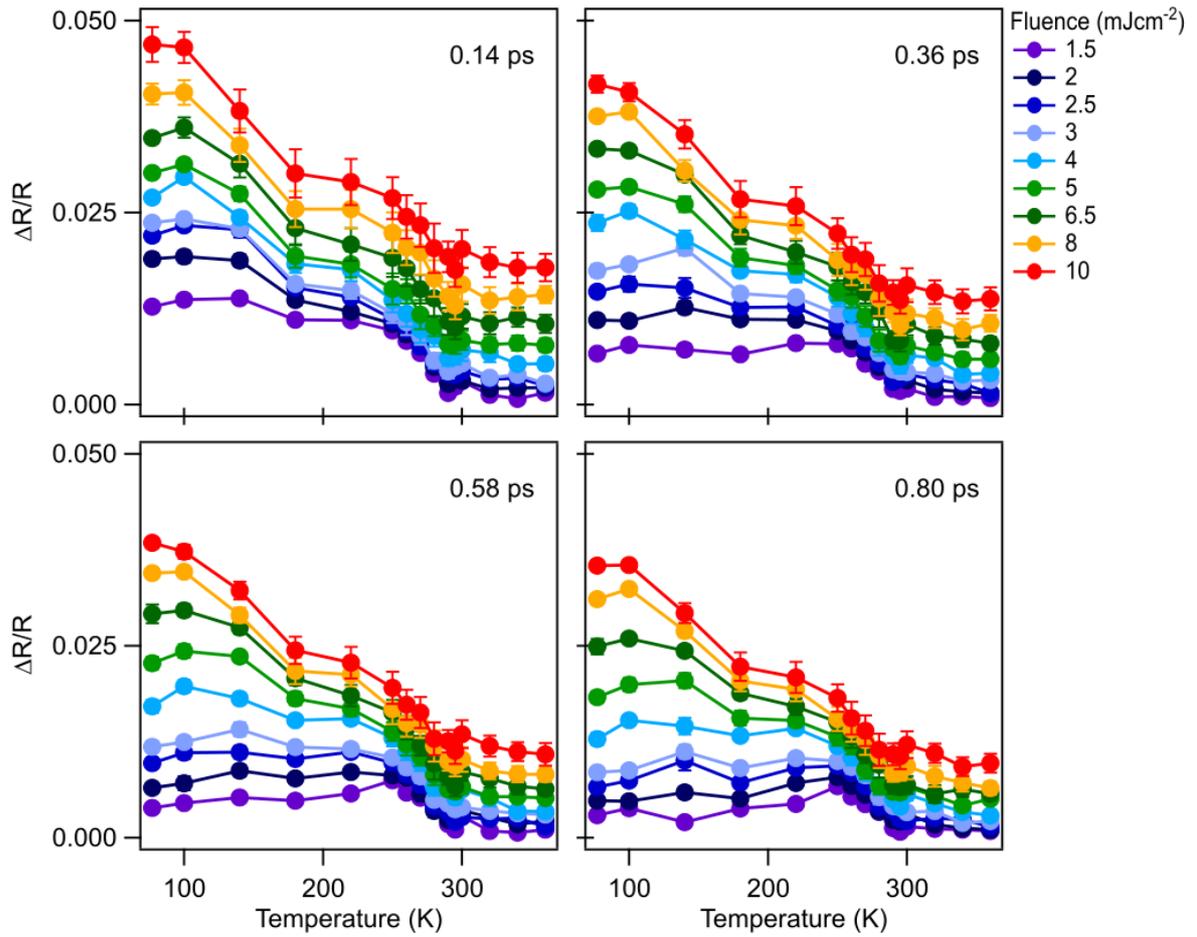

**Figure S3:** Temperature dependence of the transient reflectivity in the 610-622 nm range at different excitation fluences measured at a time delay of 0.14 ps, 0.36 ps, 0.58 ps and 0.80 ps. This wavelength is both sensitive to magnetic order while showing a temperature independent response above $T_N$. This behavior is observed at all measured fluences and time delays.

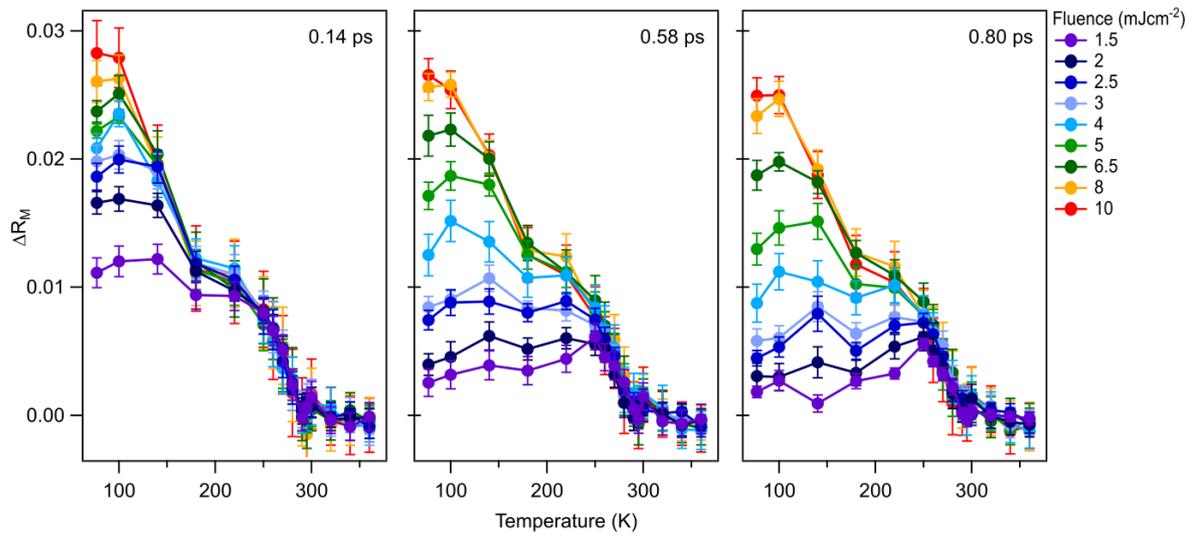

**Figure S4** Temperature dependence of the magnetic degree of freedom ($\Delta R_M$) at different excitation fluences measured at a time delay of 0.14 ps, 0.58 ps and 0.80 ps.

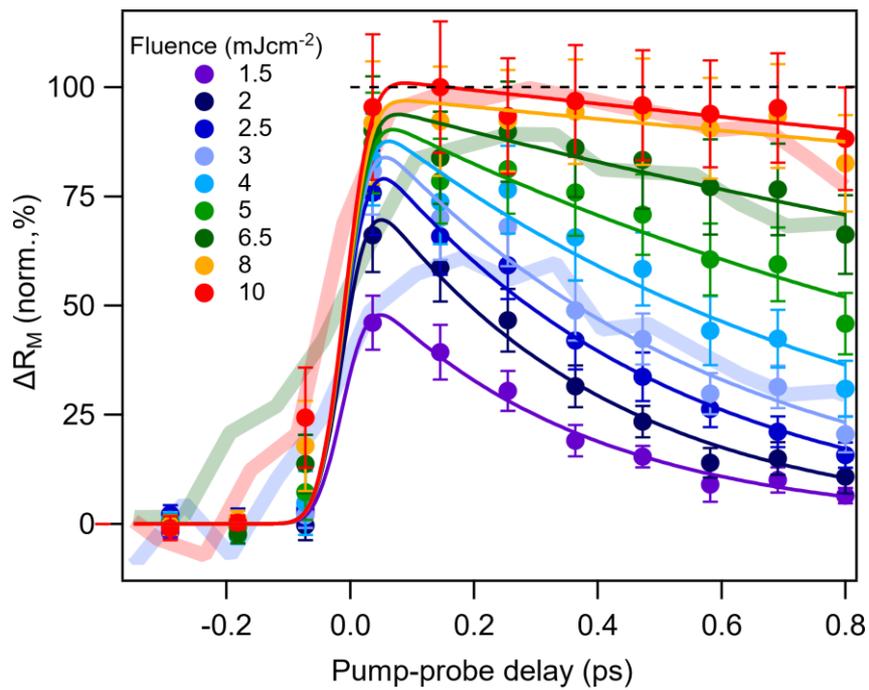

**Figure S5:** Time-evolution of the magnetic signal measured at 77 K at different fluence, extending the data in Figure 3D (main text). For comparison, the time evolution of the magnetic Bragg peak obtained from [4] is also shown (solid translucent lines, scaled).

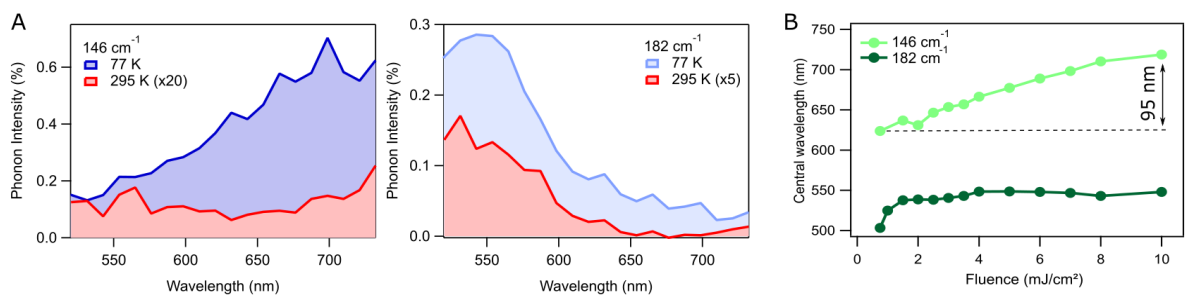

**Figure S6:** (A) Comparison of the spectral response of the two $A_{1g}$ mode at 8 mJcm$^{-2}$ as a function of temperature. The spectral dependence of the ~146 cm$^{-1}$ mode at 77 K has a Gaussian shape centered at 700 nm while at 295 K the spectra becomes broader and featureless. In contrast, the spectral shape of the ~182 cm$^{-1}$ mode is similar above and below Néel temperature with a reduced amplitude. For clarity, due to the amplitude differences, the spectral response intensity of both modes at 295 K has been multiplied by a constant factor. (B) Shift of the central frequency of the phonon spectra at 77 K. The ~146 cm$^{-1}$ frequency mode shifts by ~100 nm within 1 to 10 mJcm$^{-2}$ while the central wavelength of the ~182 cm$^{-1}$ frequency mode shows no significant change in the same fluence region.

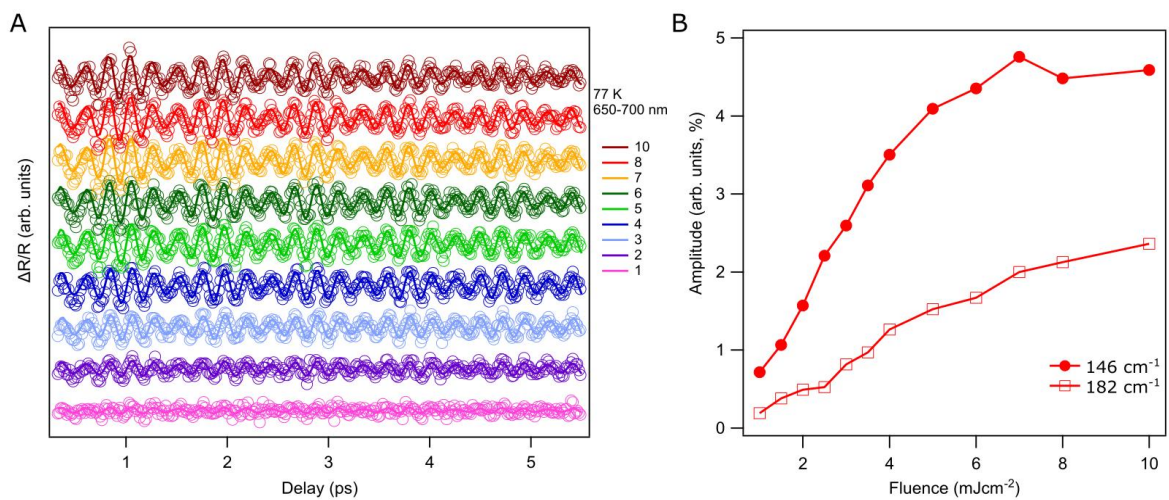

**Figure S7:** Fluence dependence on the phonon modes. (A) Fit of the differentiated transient data at 77 K and different fluence at 650-700 nm. The circles represent the raw data while the solid lines are the corresponding fit. (B) Amplitude parameter of ~146 cm$^{-1}$ and ~182 cm$^{-1}$ modes as a function of fluence obtained through the fitting procedure.

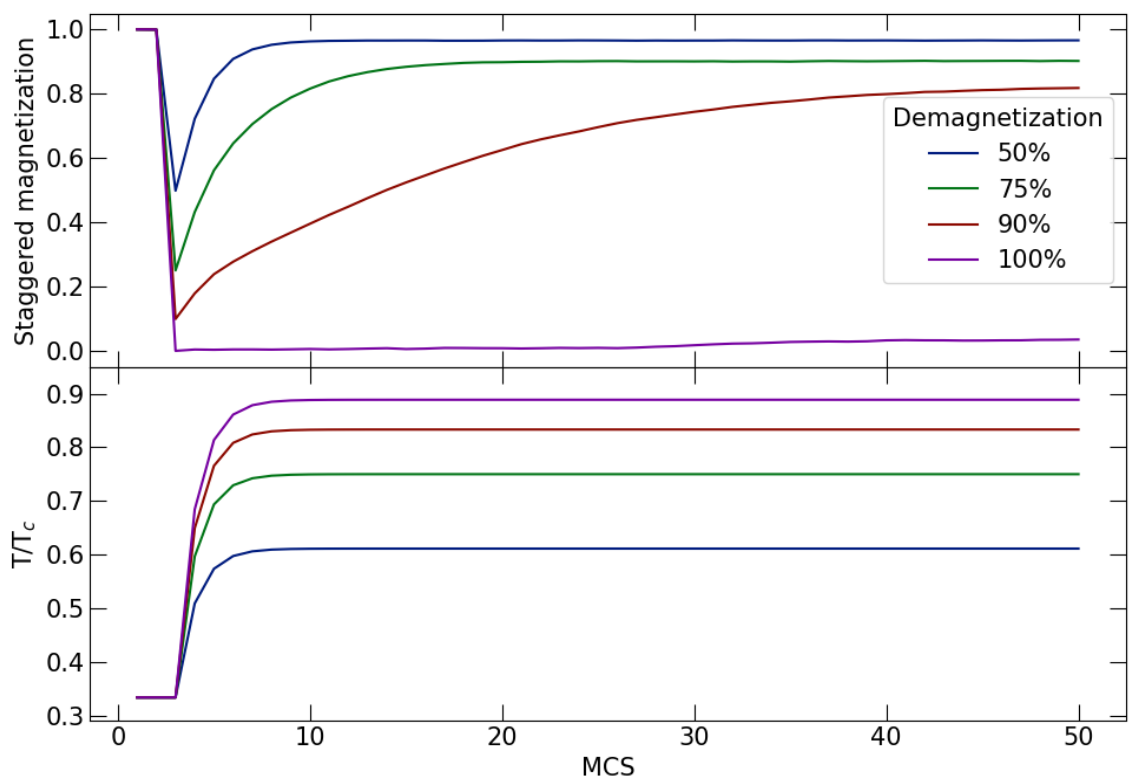

**Figure S8** Magnetization dynamics (top) and temperature profile (bottom) of a Monte Carlo 3D spin-1/2 Ising model. Details on the simulation are discussed in the text.

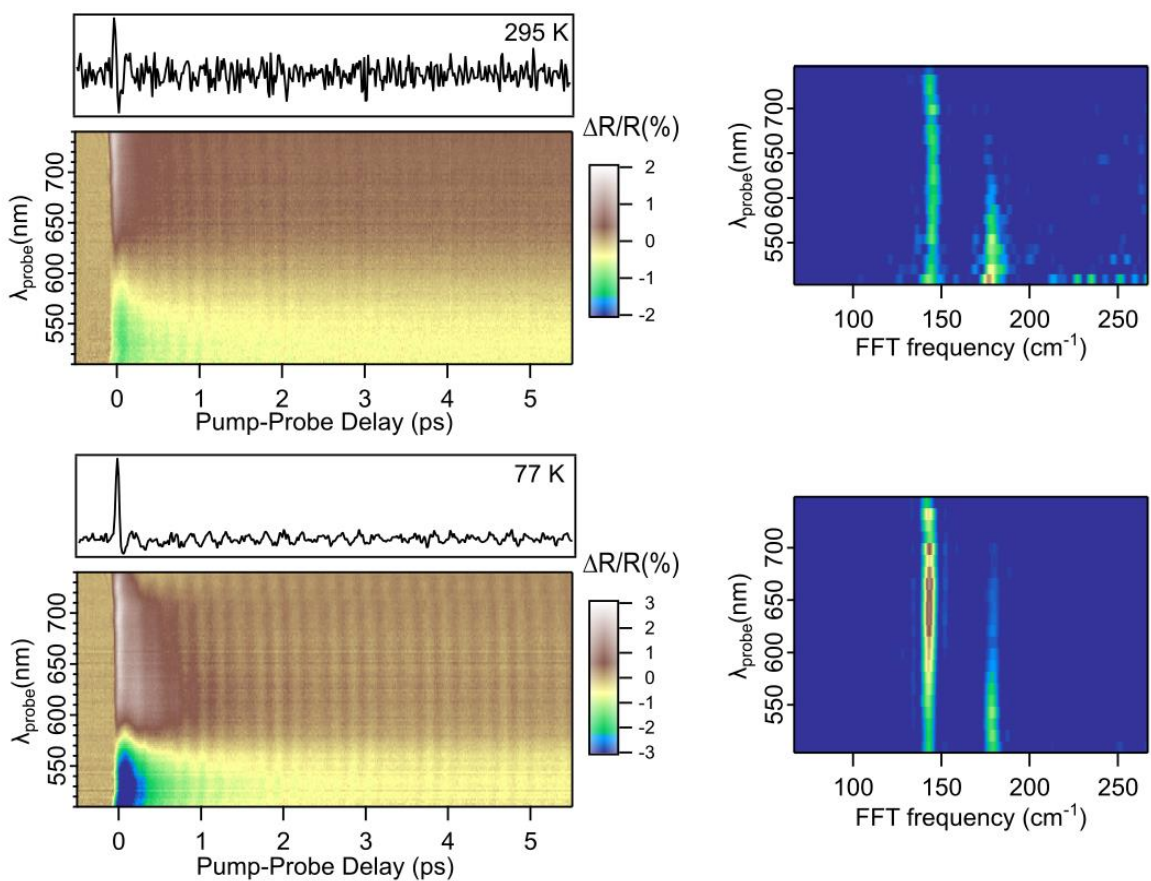

**Figure S9** (A) Differentiation of the transient reflectivity change upon 800 nm excitation ($F$= 2.5 mJ cm$^{-2}$) at 295 K and 77 K. (B) Fourier transform of the oscillations in the transient reflectivity at 295 K and 77 K for each wavelength.